\documentclass[prd,preprint,noshowpacs,tightenlines,eqsecnum,superscriptaddress,floatfix]{revtex4-1}
\usepackage{amsmath}
\usepackage{amssymb}
\usepackage{latexsym}
\usepackage[dvipsnames]{xcolor}
\usepackage{amsfonts}
\usepackage{amsthm}
\usepackage{latexsym}
\usepackage{enumitem}
\usepackage{graphics}
\usepackage{bbm}
\usepackage{tensor}
\usepackage{epstopdf}
\usepackage{hyperref}
\usepackage{graphicx}
\usepackage{caption}
\usepackage{subcaption}
\usepackage{float}
\newcommand{\be}{\begin{equation}}
\newcommand{\ee}{\end{equation}}
\newcommand{\ben}{\begin{equation*}}
\newcommand{\een}{\end{equation*}}
\newcommand{\bean}{\begin{eqnarray*}}
\newcommand{\eean}{\end{eqnarray*}}
\def\bal#1\eal{\begin{align}#1\end{align}}

\newcommand{\bsub}{\begin{subequations}}
\newcommand{\esub}{\end{subequations}}

\newcommand{\disfrac}[1][2]{\displaystyle\frac}

\newcommand*\xbar[1]{%
  \hbox{%
    \vbox{%
      \hrule height 0.5pt % The actual bar
      \kern0.5ex%         % Distance between bar and symbol
      \hbox{%
        \kern-0.1em%      % Shortening on the left side
        \ensuremath{#1}%
        \kern-0.1em%      % Shortening on the right side
      }%
    }%
  }%
}

\begin{document}
\title{General Analytic Solutions of Scalar Field Cosmology with Arbitrary Potential}
\author{N. Dimakis}
\email{nsdimakis@gmail.com}
\affiliation{Instituto de Ciencias Fisicas y Matematicas,\\
Universidad Austral de Chile, Valdivia, Chile}
\author{A. Karagiorgos}
\email{alexkarag@phys.uoa.gr}
\affiliation{Nuclear and Particle Physics Section, Physics
Department,\\
University of Athens, GR 15771 Athens, Greece}
\author{Adamantia Zampeli}
\email{azampeli@phys.uoa.gr}
\affiliation{Nuclear and Particle Physics Section, Physics
Department,\\
University of Athens, GR 15771 Athens, Greece}
\author{Andronikos Paliathanasis}
\email{anpaliat@phys.uoa.gr}
\affiliation{Instituto de Ciencias Fisicas y Matematicas,\\
Universidad Austral de Chile, Valdivia, Chile}
\author{T. Christodoulakis}
\email{tchris@phys.uoa.gr}
\affiliation{Nuclear and Particle Physics Section, Physics
Department,\\
University of Athens, GR 15771 Athens, Greece}\author{Petros A. Terzis}
\email{pterzis@phys.uoa.gr}
\affiliation{Nuclear and Particle Physics Section, Physics
Department,\\
University of Athens, GR 15771 Athens, Greece}

\begin{abstract}
We present the solution space for the case of a minimally coupled scalar field with arbitrary potential in a FLRW metric. This is made possible due to the existence of a nonlocal integral of motion corresponding to the conformal Killing field of the two-dimensional minisuperspace metric. The case for both spatially flat and non flat are studied first in the presence of only the scalar field and subsequently with the addition of non interacting perfect fluids. It is verified that this addition does not change the general form of the solution, but only the particular expressions of the scalar field and the potential. The results are applied in the case of parametric dark energy models where we derive the scalar field equivalence solution for some proposed models in the literature.
\end{abstract}
%\pacs{}

\maketitle

\section{Introduction}
The search for particular solutions to Einstein's General Relativity has been both intensive and fruitful over the past 100 years. As the years were passing, less and less simple form solutions were available to be found. This occurrence is very well depicted in the case of cosmology: for many decades, the only known solutions were simplified spatially homogeneous models; the only solutions that were subsequently recognized as the most general, were the Bianchi Type I (Kasner 1921 \cite{kasner1921geometrical}), Type II (Taub 1951 \cite{taub1951empty}) and  Type V (Joseph 1966 \cite{PSP:2056408}).
Subsequently, the automorphisms of the various Bianchi type groups were used in order to count the number of expected essential constants (Ellis and MacCallum 1969 \cite{ellis1969}), while later on automorphisms were seen to be induced by particular spacetime coordinate transformations  (Samuel and Ashtekar 1991 \cite{ashtekar1991bianchi}, Christodoulakis et al. \cite{chriskof2001}). Finally, the automorphisms were used as Lie point symmetries of the corresponding Einstein equations with the result of uncovering the entire solution space for Bianchi Types I-VII
\cite{Christodoulakis:2004yx,Christodoulakis:2006vi,Terzis:2008ev,Terzis:2010dk}.
 The situation as described above naturally led many people working in the field to turn to extended or alternative theories of gravity. It is fair to say that such novelties are supported by recent cosmological data
 (for instance \cite%
{data1,data2,data3}).

The cosmological constant $\Lambda ,~$leading to the $%
\Lambda $CDM cosmology, is one of the simplest extensions of the
Einstein-Hilbert action since it keeps linear the gravitational action, the
degrees of freedom and the order of the gravitational theory. However while
the $\Lambda $CDM cosmological model fits some of the cosmological data, it suffers from two major drawbacks, i.e. the fine tuning and the coincidence problem (for details see \cite{wein1,ccpand}).

In order to overpass these problems other theoretical models which include new
matter sources or higher-order curvature invariants in the gravitational
action have been proposed (for instance see \cite%
{BransD,faraonibook,Ame10,CapFi,odin1} and references therein). More specifically,
in scalar field cosmology, the new terms added in the gravitational Lagrangian
increase the number of the degrees of freedom, by one for one scalar field,
and the terms in the gravitational field equations are the
components of an energy-momentum tensor. The scalar field models are categorized
in two classes, according to whether the basic fields are defined in the coordinate (Einstein) frame or in its conformally equivalent (Jordan frame); in the latter there exists a coupling term in the action of
the scalar field with the curvature, for instance the Brans-Dicke
theory. Moreover other modified theories can be seen with the use of a
Lagrange multiplier as scalar-field theories, such as the $f\left( R\right) $%
-gravity in the metric formalism, the $f\left( \mathcal{R}\right) $-hybrid
gravity (where $ \mathcal{R}$ is the scalar curvature in the Palatini formalism and coincides with the Ricci scalar only for a linear function $f$), or the higher-order $f\left( R,\square R,..\right) $-gravity \cite%
{Wands1,CapUni,GaoPLB}. However, under conformal transformations the two
different frames, Einstein and Jordan, are related which means that a
cosmological solution can pass from the one frame to the other \cite%
{frame1,Kaiser1}.

In this work we follow the old path focusing attention to an isotropic and spatially homogeneous model with matter and directing our analysis  towards the investigation of the entire solution space. Specifically, we consider a  Friedmann-Lema\^{\i}%
tre-Robertson-Walker (FLRW) spacetime and a minimally coupled scalar field with arbitrary potential (for relevant results see \cite{christodoulakis2006decoupling}, in which the corresponding to the automorphisms local integrals of motion are used to completely integrate the relevant equations for the Bianchi Type I and V diagonal metrics; thus, the $k=0, \, -1$ cases of the present work should somehow be included in this reference, if only one could find the relevant transformation). We are interested in the analytic
solutions of that cosmological model. Exact closed-form solutions without
matter term or with a dust fluid can be found in ~\cite{muslimov1990scalar,ellis1991exact,barrow1,mendez1996exact,russo2004exact,basil1,palia1,ester12,Fre,palia2,sup16} and in the case of a non-minimal coupling in \cite{Vernov1,Vernov2,Vernov3}, while for non-spatially flat FLRW spacetimes some exact solutions are contained in  \cite{halliwell1987scalar,ea1993,Gum08}. Further analytic
solutions for a scalar field model with the presence of a perfect fluid have
been presented in \cite{chimento1996r,urena00,star11,barow2,palia3}.

The major ingredient which enables the exhibition of the entire solutions space of our model is the use of nonlocal conservation laws, which are generated
by the elements of the minisuperspace conformal algebra. The general analytic solution is presented for every arbitrary scalar field potential $V(\phi)$, for a quintessence or for a phantom field in both  spatially flat and/or non-flat FLRW  metric. Moreover, we show that the same result also
holds when additional non interacting perfect fluids are introduced in the field equations.

The plan of the paper is as follows.
The basic definitions for our model as also the mathematical properties that
we use are discussed in Section \ref{basic}. In Sections \ref{flat} and \ref%
{nonflat} we derive the analytic solution for a minimally coupled scalar
field in a FLRW spacetime for arbitrary
potential for the cases in which the spatial curvature is $k=0$, and $k\neq 0
$, respectively. In Section \ref{perfect} we consider that our cosmological
model admits perfect fluids which are minimally coupled with the scalar
field and we derive the analytic solution of the model for arbitrary
potential. Furthermore, in section \ref{examples} we demonstrate the usefulness and analytical power of our results, by deriving some particular solutions of the field equations for
specific equation of state parameters of the total cosmological fluid. In
section \ref{conclusions} we discuss our results and we draw our conclusions.

\section{Basic Definitions \label{basic}}

It is a known fact that for many cosmological systems there is a procedure
of deriving valid mini-superspace Lagrangians, whose dynamical content is
the same with that of the original system. These Lagrangians are - by
construction - singular in nature, since their equations of motion are not
all independent of each other (as is also true for the general theory). This
property is reflected in the time reparameterization invariance of the
system $t\mapsto f(t)$, which is a remnant of the four dimensional
diffeomorphism invariance of the full theory.

It was proven in \cite{tchris1}, that for singular systems described by
Lagrangians of the form
\begin{equation}
\xbar{L}=\frac{1}{2N(t)}\xbar{G}_{\mu \nu }(q)\dot{q}^{\mu }(t)\dot{q}^{\nu
}(t)-N(t)\,U(q)  \label{general_lag}
\end{equation}%
all conformal Killing fields of $\xbar{G}_{\mu \nu }$ can be used to write
down integrals of motion for the system. To make it explicit, let us for
simplicity perform a reparameterization of the form $N\rightarrow n=N\,U$.
Then, it can be seen that for the equivalent system
\begin{equation}
L=\frac{1}{2n}G_{\mu \nu }(q)\dot{q}^{\mu }\dot{q}^{\nu }-n,
\label{general_lag_const}
\end{equation}%
where $G_{\mu \nu }=U\,\xbar{G}_{\mu \nu }$, if there exist vector fields $%
\xi ^{\alpha }(q)$ defined on the configuration space for which the relation
\begin{equation}
\emph{\textsterling}_{\xi }G_{\mu \nu }=\omega (q)G_{\mu \nu }
\label{general_conf}
\end{equation}%
holds, then the quantity
\begin{equation}
Q=\xi ^{\alpha }p_{\alpha }+\int \!\!n(t)\omega (q(t))dt,
\label{general_int}
\end{equation}%
with $p_{\alpha }=\frac{\partial L}{\partial q^{\alpha }}$, defines
integrals of motion on the phase space due to the existence of the
Hamiltonian constraint
\begin{equation}
\mathcal{H}=\frac{1}{2}G^{\mu \nu }p_{\mu }p_{\nu }+1\approx 0
\label{general_Ham}
\end{equation}%
It can be easily verified, that by virtue of \eqref{general_conf}, $Q$ is a
constant of motion
\begin{equation*}
\frac{dQ}{dt}=\frac{\partial Q}{\partial t}+\{Q,\mathcal{H}\}=\omega
\mathcal{H}\approx 0.
\end{equation*}

In the special case were $\omega =0$, i.e. the $\xi$'s are Killing vectors
of the scaled mini-supermetric $G_{\mu\nu}$, relation \eqref{general_int}
leads to autonomous integrals of motion that do not exhibit any explicit
time dependence. On the other hand, when $\omega \neq 0$, nonlocal
integrals of motion emerge that are actually rheonomic, due to the explicit
time dependence brought by the integral in \eqref{general_int}.

The autonomous conserved quantities are commonly used in the bibliography as
a kind of selection rule to constrain the potential $U$, or any arbitrary functions
of the configuration variables appearing in \eqref{general_lag}, so that the
system is forced to become integrable \cite{palia1,palia2,palia3,Babak,Dimakis:2013oza,paterz,palia4,palia5}. In our case, we shall refrain from doing that.
In the context of Einstein's relativity and a spatially flat/non-flat FLRW
spacetime minimally coupled with a scalar field $\phi$, we use the most
general rheonomic integrals of motion we can write down, so that the
equations of motion can be solved for any arbitrary potential $V(\phi)$. The
importance of this fact is twofold: not only it provides a mapping between
metrics and scalar field potentials for which the former are solutions to
Einstein's equations, but also proves that for this particular configuration
(FLRW plus minimally coupled scalar field) the induced system is completely
integrable for any (smooth enough) function $V(\phi)$.

The action of the system under consideration is
\begin{equation}  \label{act}
S = \int \!\! d^4 x \sqrt{-g} \left(R + \epsilon \, \phi_{,\mu} \phi^{,\mu}
+ 2\, V(\phi)\right)
\end{equation}
where $R$ is the Ricci scalar and $g$ the determinant of the spacetime
metric $g_{\mu\nu}$. We also allow for the existence of a phantom field by
introducing a pure sign constant $\epsilon = \pm 1$. As it is well known, variation of the above action
with respect to $g^{\mu\nu}$ yields the equations
\begin{equation}  \label{ein}
R_{\mu\nu} - \frac{1}{2} R g_{\mu\nu} = T_{\mu\nu}
\end{equation}
with
\begin{equation}  \label{energy_mom}
T_{\mu\nu} = \epsilon\, \phi_{,\mu} \phi_{,\nu} - \frac{1}{2}
\left(\epsilon\, \phi^{,\kappa}\phi_{,\kappa} -2\, V(\phi) \right) g_{\mu\nu}
\end{equation}
being the energy-momentum tensor for the matter part. Variation of %
\eqref{act} with respect to the scalar field leads to the Klein-Gordon equation (with arbitrary potential)
\begin{equation}  \label{KGeq}
\epsilon\, \square \phi - V^{\prime }(\phi) = 0,
\end{equation}
where the prime denotes differentiation with respect to the field $\phi$.

It can be easily shown that the assumption of the  following form for the metric
\begin{equation}  \label{FLRW}
ds^2 = - N(t)^2 dt^2 + a(t)^2 \left(\frac{1}{1-k\, r^2} dr^2 + r^2 d\theta^2 +
r^2 \sin^2\theta d\varphi^2 \right)
\end{equation}
leads to the necessity for the scalar field to be spatially homogeneous i.e $\phi = \phi(t)$.
The set of equations (\ref{ein}-\ref{KGeq}), when reduced by the above demands, is equivalent to
the Euler-Lagrange equations derived by the Lagrangian
\begin{equation}  \label{miniLag}
L = \frac{2 a^2}{n}\left(a^2 V(\phi)-3 k\right) \left( - 6 \dot{a}^2 +
\epsilon \, a^2 \dot{\phi}^2\right) - n,
\end{equation}
with the dots indicating the time derivatives. Note that we have written the
Lagrangian in the form \eqref{general_lag_const}, the scaled lapse function $%
n$ is related to the original $N$ appearing in \eqref{FLRW} by
\begin{equation}  \label{lapserel}
N = \frac{n}{2\, a \left(a^2 V(\phi) - 3 k\right)}.
\end{equation}
We choose to start working in the lapse parametrization which leads to a
Lagrangian with a constant potential $U(q) =1$, due to the fact that it
renders the relations for the corresponding integrals of motion simpler. The
scaled mini-supermetric in this case is
\begin{equation}  \label{minisup}
G_{\mu\nu} =4\, a^2 \left(a^2 V(\phi)-3 k\right)
\begin{pmatrix}
- 6 & 0 \\
0 & \epsilon\, a^2%
\end{pmatrix}%
\end{equation}
and any of its conformal Killing fields can be used to define integrals of
motion for the corresponding system.

\section{Spatially flat FLRW spacetime\label{flat}}

Although - as we shall see in the next section - the cases $k=0$ and $k\neq
0 $ can be treated simultaneously, we choose to express the solution for the
spatially flat case separately, since its simplicity helps to a better
understanding of the implemented methodology.

In the $k=0$ case, the mini-supermetric \eqref{minisup} exhibits a
homothetic vector
\begin{equation}  \label{hom}
\xi = \frac{a}{6} \frac{\partial}{\partial a}
\end{equation}
which is independent of the scalar field potential $V(\phi)$ and satisfies $\emph{\textsterling}_\xi G_{\mu\nu} = G_{\mu\nu}$. This
results in the existence of a conserved quantity in phase space that is
written as
\begin{equation}  \label{Qh}
\begin{split}
Q & = \frac{a}{6} p_a + \int\!\! n(t) dt = \frac{a}{6} \frac{\partial L}{%
\partial \dot{a}} + \int\!\! n(t) dt \\
& = -\frac{4\, a^5 \dot{a} V(\phi)}{n} + \int\!\! n(t) dt .
\end{split}%
\end{equation}
Thus, we are led to consider the relation
\begin{equation}  \label{int_eq}
Q = \kappa,
\end{equation}
with $\kappa$ a constant, as a first integral for the relevant system of
equations of motion.

Before proceeding let us mention a few facts from the theory of constrained
systems. The number of true degrees of freedom is found by the relation $%
\frac{1}{2} (M-2F-S)$, where $M$ is the dimension of the full phase space,
while $F$ and $S$ are the numbers of first and second class constraints
respectively. In our case the full phase space is spanned by $n$, $a$, $\phi$
together with the corresponding momenta. At the same time there exist two
first class constraints $p_n \approx 0$ and $\mathcal{H} \approx 0$, both representing the invariance of the action under arbitrary time reparametrizations $t=f(\tilde{t})$; which
means that there exists only one true degree of freedom. This can be seen by algebraically solving the constraint equation $\frac{\partial L}{\partial n}=0$ with respect to $n$ and substituting the result into the two remaining Euler-Lagrange equations. Then, it is found that only one independent equation remains. Thus, the general solution can be obtained by solving any convenient combination of $t=f(a, \phi)$ as the time parameter.
This interesting property of constrained systems can be exploited for adopting
different and more convenient gauge choices, in order for the system of
equations to be integrated. In what follows we appoint $\phi$ as the time
variable $t$, while $n$ and $a$ are to be derived with the help of the
integrals of motion and one of the Euler-Lagrange equations. Of course, the
choice $\phi =t$ has as a prerequisite the assumption that $\phi$ can be
considered - at least locally - as an invertible function of time. The
latter is guaranteed by the inverse function theorem as long as $\phi(t)$ is
differentiable with $\dot{\phi} \neq 0$. As a result, the case of a constant
scalar field - which corresponds to the known pure cosmological constant
solution - cannot be included in the following analysis. Of course, it can be considered as a separate case, for $\phi =c$ and by selecting the scale factor $a$ as the time parameter. Let us now proceed by fixing the gauge through the choice $\phi(t) =
t$. At the same time we express the scaled lapse function $n(t)$ with the
help of a new non-constant function $h(t)$ and we reparameterize the
potential - that now can be considered as a function of time $V(t)$ - with
respect to a new (again non-constant) function $A(t)$:
\begin{subequations}
\label{reparam1}
\begin{align}  \label{reparam1V}
n(t) & = \dot{h}(t), \\
V(t) & = \frac{\left(h(t) - \kappa \right) \dot{h}(t)}{4 \dot{A}(t)} .
\end{align}
As a result, \eqref{int_eq} reduces to a local expression
\end{subequations}
\begin{equation*}
a(t)^5 \dot{a}(t)- \dot{A}(t) =0,
\end{equation*}
that can be easily integrated to give
\begin{equation}  \label{sola1}
a(t) = \pm 6^{1/6} \left(A(t)+c_1 \right)^{1/6}.
\end{equation}
Substitution of the above solution into the Euler-Lagrange equation for $%
a(t)$ leads to
\begin{equation}  \label{eula1}
2 (A(t)+c_1) \dot{A}(t) \dot{h}(t)+(\kappa - h(t)) \left(\dot{A}(t)^2-6\,
\epsilon \, (A(t)+c_1)^2\right) =0,
\end{equation}
which implies that
\begin{equation}  \label{solA1}
A(t) = \frac{\mu^4}{6}\exp \left(- \int\! \frac{\dot{h} \pm \left(\dot{h}^2
+ \epsilon \, (\kappa -h)^2 \right)^{1/2}}{\kappa - h} \, dt \right) - c_1 ,
\end{equation}
with $\mu$ a non zero constant. It can be easily checked that %
\eqref{reparam1}, together with \eqref{sola1} and \eqref{solA1} completely
solve the system of the Euler-Lagrange equations of the Lagrangian %
\eqref{miniLag} for $k=0$ in the gauge $\phi=t$ (and hence Einstein's
equations \eqref{ein}). The function $h(t)$ remains free, reflecting the
arbitrariness of the potential $V(t)$ through \eqref{reparam1V}.

It can be seen by \eqref{sola1} and \eqref{solA1}, that the constant $c_1$
is not important for the solution, so we might as well consider it to be
zero. The same is true for $\kappa$, since one needs only define a new
parametrization as $h(t) = \kappa + \exp\left(\frac{\omega}{2}-3 \int\!
\frac{\epsilon}{\dot{\omega}} dt\right)$. It can be verified that in this form for $h(t)$, the lapse function $N(t)$, as given by \eqref{lapserel},
together with the scale factor $a(t)$ and the potential $V(t)$ assume the
values
\begin{equation}  \label{dotompos}
N(t) = \frac{1}{3} \mu ^2 \dot{\omega}\, e^{3 \int\! \frac{\epsilon}{\dot{%
\omega}} dt}, \quad a(t) = \mu^{2/3} e^{\omega/6}, \quad V(t) = \frac{3
\left(\dot{\omega}^2-6\, \epsilon \right) e^{-6 \int\! \frac{\epsilon}{\dot{%
\omega}} dt}}{4\, \mu ^4 \dot{\omega}^2}
\end{equation}
when $\dot{\omega}>0$ and
\begin{equation}  \label{dotomneg}
N(t) = -\frac{2\, \epsilon \, \mu ^2 e^{-\frac{\omega}{2}}}{\dot{\omega}},
\quad a(t) = \mu^{2/3} e^{-\int \frac{\epsilon}{\dot{\omega}} dt}, \quad
V(t) = -\frac{e^{\omega} \left(\dot{\omega}^2-6\, \epsilon \right)}{8\,
\epsilon\, \mu^4}
\end{equation}
when $\dot{\omega}<0$ (we have chosen to adopt the plus solution in %
\eqref{solA1}, the same apply for the minus case with an interchange of the
previous relations with respect to the sign of $\dot{\omega}$). However, it
is easy to check that in the second case of \eqref{dotomneg}, if we choose
to express $\omega$ with respect to a new function $g(t)$, through a
relation $\omega = -6 \int\! \frac{\epsilon}{\dot{g}} dt$, then we acquire
the first set of relations \eqref{dotompos} with $g(t)$ in place of $%
\omega(t)$. Henceforth, given the fact that $\omega$ in the solution is
arbitrary, without loss of generality we can consider only \eqref{dotompos},
since the solution can always be brought into this form. The resulting line
element with the additional help of a scaling $r \mapsto r \mu^{-2/3}$
and a reparametrization $\mu=\sqrt{3}\, m$ can be written as
\begin{equation}  \label{lineel1}
ds^2 = -m^4 \dot{\omega}^2 e^{6 \int\! (\epsilon/\dot{\omega}) dt} dt^2 +
e^{\omega/3} \left( dr^2 +r^2 d\theta^2 + r^2 \sin^2 \theta d\varphi^2
\right) .
\end{equation}
and the respective scalar field potential for each $\omega(t)$ is
\begin{equation}  \label{potential1}
V(t) = \frac{ \left(\dot{\omega}^2-6\, \epsilon \right) e^{-6\int\!
(\epsilon/\dot{\omega}) dt}}{12 m^4 \dot{\omega}^2}.
\end{equation}
So, given any non-constant function $\omega$, there is a line-element %
\eqref{lineel1} satisfying the equations of motion with the corresponding
potential \eqref{potential1}.

Solution \eqref{lineel1} can be further simplified by performing a change in
the time variable from $t$ to $\omega $. Since, $\omega $ is an arbitrary
function of $t$, we choose to invert the relation $\omega (t)$ by the use of
an arbitrary function $F(\omega )$ defined as follows:
\begin{equation}
t=\int \!\!\sqrt{\epsilon \frac{F^{\prime }(\omega )}{6}}d\omega ,
\label{omegatoF}
\end{equation}%
the prime denoting differentiation with respect to the argument $\omega $.
The line element \eqref{lineel1} - with a slight redefinition of the $%
F(\omega )$ function ($F(\omega )\mapsto F(\omega )-\log m^{4}$) that
does not alter \eqref{omegatoF} - can be written
\begin{equation}
ds^{2}=-e^{F(\omega )}d\omega ^{2}+e^{\omega /3}\left( dr^{2}+r^{2}d\theta
^{2}+r^{2}\sin ^{2}\theta d\varphi ^{2}\right)  \label{lineel1b}
\end{equation}%
while the potential \eqref{potential1} transforms to
\begin{equation}
V(\omega )=\frac{1}{12}e^{-F(\omega )}\left( 1-F^{\prime }(\omega )\right) .
\label{potential1b}
\end{equation}%
and of course the scalar field that completes the solution is
\begin{equation}
\phi (t)=t\mapsto \phi (\omega )=\int \!\!\sqrt{\epsilon \frac{F^{\prime
}(\omega )}{6}}d\omega .  \label{phi1b}
\end{equation}%
The set (\ref{lineel1b}-\ref{phi1b}) satisfies
the Einstein plus Klein-Gordon equations in the new time variable $\omega $,
with $F(\omega )$ remaining of course an arbitrary function due to the fact
that we have not adopted a particular form for the potential. It is to be
noted, that exactly the same procedure applies if instead of \eqref{omegatoF}%
, we consider the time change $t=-\int \!\!\sqrt{\epsilon \frac{F^{\prime
}(\omega )}{6}}d\omega $. The only thing that alters is the sign in front
of the integral in \eqref{phi1b}. Thus, for line element \eqref{lineel1b}
and potential \eqref{potential1b}, both $+/-$ solutions for $\phi $ are
valid.

Given the energy momentum tensor \eqref{energy_mom}, it is a well known fact
that the behaviour of matter due to the scalar field can be effectively
simulated by a perfect fluid, whose energy density and pressure are
\begin{subequations}
\label{efffluid}
\begin{align}
\rho_\phi (t) & = T^{\mu\nu} u_\mu u_\nu \\
P_\phi (t) & = \frac{1}{3} T^{\mu\nu} h_{\mu\nu}
\end{align}
where $u_\mu= \frac{\phi_{,\mu}}{\sqrt{-g^{\kappa\lambda}\phi_{,\kappa}%
\phi_{,\lambda}}}$ is the comoving four velocity and $h_{\mu\nu} =
g_{\mu\nu} + u_\mu u_\nu$ the metric of the three-surfaces normal to the
direction of $u_\mu$.

In our case, solution (\ref{lineel1b}-\ref{phi1b}) leads to the simple
expressions
\end{subequations}
\begin{subequations}
\label{rhoP1}
\begin{align}
\rho _{\phi }(\omega )& =\frac{1}{12}e^{-F(\omega )} \\
P_{\phi }(\omega )& =\frac{1}{12}e^{-F(\omega )}\left( 2F^{\prime }(\omega
)-1\right)
\end{align}%
wherefrom we can deduce the equation of state
\end{subequations}
\begin{equation}
P_{\phi }=(2F^{\prime }(\omega )-1)\rho _{\phi }.  \label{eqofst1}
\end{equation}
and we can see that the parameter $\gamma_\phi = \frac{P_\phi}{\rho_\phi}$ contains now the arbitrary function $F (\omega)$. This expression allows for the study of general cases for the scalar field potential.

\section{FLRW with spatial curvature\label{nonflat}}

Due to the fact that, when $k\neq 0$, the configuration space vector \eqref{hom} no longer generates a homothecy of $G_{\mu\nu}$, the situation becomes more
complicated. However, it can be seen that there exists a conformal vector $%
\xi = \frac{\partial}{\partial \phi}$, with a corresponding factor $\frac{%
a^2 V^{\prime }(\phi )}{a^2 V(\phi )-3 k}$, i.e.
\begin{equation}  \label{conf}
\emph{\textsterling}_\xi G_{\mu\nu} = \frac{a^2 V^{\prime }(\phi )}{a^2
V(\phi )-3 k} G_{\mu\nu}.
\end{equation}
Subsequently, the following nonlocal integral of motion can be defined
\begin{equation}  \label{Qconf}
\begin{split}
Q & = p_\phi + \int\!\! \frac{a(t)^2 n(t) V^{\prime }(\phi (t))}{a(t)^2
V(\phi (t))-3 k} dt = \frac{\partial L}{\partial \dot{\phi}} + \int\!\!
\frac{a(t)^2 n(t) V^{\prime }(\phi (t))}{a(t)^2 V(\phi (t))-3 k} dt \\
& = \frac{4\, \epsilon\, a^4 \dot{\phi} \left(a^2 V(\phi )-3 k\right)}{n} +
\int\!\! \frac{a(t)^2 n(t) V^{\prime }(\phi (t))}{a(t)^2 V(\phi (t))-3 k} dt
.
\end{split}
\end{equation}
It can be straightforwardly checked that, equation $Q = \kappa$ is a first
integral of the Klein-Gordon equation \eqref{KGeq} (or equivalently, of the
Euler-Lagrange equation with respect to $\phi$).

Again, we choose the gauge $\phi(t) = t$ and parametrize the dependent variables as
\begin{subequations}
\label{reparam2}
\begin{align}  \label{reparam2V}
n(t) & = \frac{2 \dot{h} \left(a^2 V-3 k\right)}{a^2 \dot{V}} \\
V(t) & = \int\!\! \frac{\dot{w} }{a^6} dt,
\end{align}
with $w(t)$ and $h(t)$ being non constant functions of time. Now, the
corresponding equation \eqref{int_eq} reduces to
\end{subequations}
\begin{equation*}
\frac{2\, \epsilon\, \dot{w}}{\dot{h}}+2 h-\kappa =0
\end{equation*}
and can be immediately integrated to yield
\begin{equation}  \label{solh}
h(t) = \frac{1}{2} \left(\kappa \pm \sqrt{4\, c_1 + \kappa^2 - 8\,
\epsilon\, w} \right).
\end{equation}
By choosing to express $w(t)$ as
\begin{equation}  \label{wpar}
w(t) = \frac{a \dot{v}}{\dot{a}}+\frac{1}{8\, \epsilon} \left(4\, c_1 +
\kappa ^2\right)-6 v,
\end{equation}
where $v(t)$ is a new function of time, substitution of \eqref{reparam2}
together with \eqref{solh} into the quadratic constraint $\frac{\partial L}{%
\partial n}=0$ leads to
\begin{equation}  \label{quadcon}
-\frac{6\, \dot{a}\, \dot{v}}{\epsilon\, a}+\frac{36\, v\, \dot{a}^2}{%
\epsilon\, a^2}+3\, k\, a^4-6\, v =0,
\end{equation}
with solution
\begin{equation}  \label{solu}
v(t) = \exp \left(\int\!\! \left(\frac{6 \dot{a}}{a}-\frac{\epsilon\, a}{%
\dot{a}}\right) dt\right) \left(\int\!\! \frac{k a^5 \exp \left(-\int\!\!
\left(\frac{6 \dot{a}}{a}-\frac{\epsilon\, a}{\dot{a}}\right) \, dt\right)}{%
2 \dot{a}} dt+ c_2 \right),
\end{equation}
$c_2$ being an integration constant. Thus, the process is complete and
the resulting line-element - by setting for simplicity $a(t) = e^{\omega/6}$
- can be written as
\begin{equation}  \label{lineel2}
\begin{split}
ds^2 = & \frac{- e^{\omega} \dot{\omega}^2}{36 \left( 2\, e^{\omega -6 \int
(\epsilon/\dot{\omega}) dt} \left(c_2 +3\, k \int \frac{\exp \left(6\int
(\epsilon/\dot{\omega}) dt-\frac{\omega}{3}\right)}{\dot{\omega}} dt\right)
- k e^{\frac{2 \omega}{3}}\right)} dt^2 \\
& + e^{\omega/3} \left( \frac{1}{1-k r^2} dr^2 + r^2 d\theta^2 +r^2 \sin^2
\theta d\varphi^2\right)
\end{split}%
\end{equation}
which for $k=0$ can, by an
appropriate reparametrization of integration constant $c_2= \frac{1}{72\, m^4}$, be brought exactly into the form \eqref{lineel1}. The corresponding scalar field potential for each non-constant function $%
\omega$ is
\begin{equation}  \label{potential2}
V(t) = \frac{6\, e^{-\omega} \left(\left(\dot{\omega}^2-6\, \epsilon\right)
e^{\omega -6 \int (\epsilon/\dot{\omega}) dt} \left(c_2 + 3\, k \int \frac{%
\exp \left(6 \int (\epsilon/\dot{\omega}) - \frac{\omega}{3} dt\right)}{\dot{%
\omega}} dt\right)+ 3\, k \, e^{\frac{2 \omega }{3}}\right)}{\dot{\omega}^2},
\end{equation}
which again, for $k=0$ and $c_2= \frac{1}{72\, m^4}$, becomes %
\eqref{potential1}.

By adopting a suitable time change, as in the previous section, the result
can be significantly simplified. We perform the transformation (where again
we utilize a non constant function $S(\omega )$)
\begin{equation}
t=\pm \int \!\left[ \frac{1}{6\,\epsilon }\left( \frac{S^{\prime \prime
}(\omega )}{S^{\prime }(\omega )}+\frac{1}{3}\right) \right]
^{1/2}\!\!d\omega  \label{omegatoF2}
\end{equation}%
(for any of the two signs in the above equation the treatment is exactly the
same). Then, line element \eqref{lineel2} - with the help of an allowable ($k \neq 0$) redefinition
\begin{equation}
S(\omega )=\exp \left( 12\,k\int \!e^{F(\omega )-\omega /3}d\omega \right) -%
\frac{6\,c_{2}}{k}  \label{StoF}
\end{equation}%
which leads to the absorption of the non-essential constant $c_{2}$ - simplifies to
\begin{equation}
ds^{2}=-e^{F(\omega )}d\omega ^{2}+e^{\omega /3}\left( \frac{1}{1-kr^{2}}%
dr^{2}+r^{2}d\theta ^{2}+r^{2}\sin^2 \theta d\varphi ^{2}\right) ,
\label{lineel2b}
\end{equation}%
while the corresponding potential becomes
\begin{equation}
V(\omega )=\frac{1}{12}e^{-F(\omega )}\left( 1-F^{\prime }(\omega )\right)
+2\,k\,e^{-\omega /3}  \label{potential2b}
\end{equation}%
with the scalar field $\phi (\omega )$ - since in the previous gauge $\phi
(t)=t$ - given now by \eqref{omegatoF2} after substitution of \eqref{StoF}
\begin{equation}
\phi(\omega) = \pm \int\!\! \left[ \frac{1}{6\, \epsilon } \left(F'(\omega)+12\, k\, e^{F(\omega)-\omega/3}\right) \right]^{1/2} \! d\omega ,  \label{phi2b}
\end{equation}%
where the $+$ or $-$ sign corresponds to the relevant choice of time
transformation in \eqref{omegatoF2}.

As a result, the set of relations (\ref{lineel2b}-\ref{phi2b}) satisfies
Einstein's equation plus scalar field for the case $k\neq 0$, with $%
F(\omega) $ remaining again arbitrary. It can be seen that if one considers $%
k=0$, then solution (\ref{lineel2b}-\ref{phi2b}) becomes exactly %
\eqref{lineel1b} - \eqref{phi1b}. It is noteworthy that this occurs despite the fact that in the
process of deriving relations (\ref{lineel2b}-\ref{phi2b}) the
assumption $k\neq 0$ has been taken into account (see \ref{StoF}).

As we did in the previous case, we can - starting from the relations \eqref{efffluid} -
compute the energy density and the pressure of the matter content
in terms of the function $F(\omega )$
\begin{subequations}
\label{rhoP2}
\begin{align}
\rho _{\phi }(\omega )& =\frac{1}{12}e^{-F(\omega )}+3\,k\,e^{-\omega /3} \\
P_{\phi }(\omega )& =\frac{1}{12}e^{-F(\omega )}\left( 2F^{\prime }(\omega
)-1\right) -k\,e^{-\omega /3},
\end{align}%
leading to the equation of state
\end{subequations}
\begin{equation}
P_{\phi }=\left( \frac{2\, e^{\omega /3}\left( 3F^{\prime }(\omega )-1\right) }{%
3\left( 36\,k\,e^{F(\omega )}+e^{\omega /3}\right) }-\frac{1}{3}\right) \rho
_{\phi }.  \label{eqofst2}
\end{equation}%
Again we can notice the difference with respect to \eqref{rhoP1} due to the
contribution of $k$.

\section{Inclusion of an additional perfect fluid\label{perfect}}

All the previous considerations can be slightly modified to consider an
additional perfect fluid matter source together with the scalar field. The
extra contribution to the energy momentum tensor is given by
\begin{equation}  \label{perffluidem}
\mathcal{T}_{\mu\nu} = (\rho + P) \tilde{u}_\mu \tilde{u}_\nu + P g_{\mu\nu},
\end{equation}
where $\tilde{u}^\mu= (1/N(t),0,0,0)$ is the four velocity of the co-moving
observer and $\rho$, $P$ the energy density and the pressure of the fluid
respectively. In what follows we shall assume a barotropic equation of state
of the form $P= \gamma \rho$, with $\gamma$ being a constant. It is known,
that the energy momentum tensor of a perfect fluid \eqref{perffluidem} can
be recovered by varying the matter Lagrangian density $\mathcal{L}_m \propto
\rho$ with respect to $g_{\mu\nu}$ when a continuity equation $\mathcal{T}%
^{\mu\nu}_{\phantom{\mu\nu} ;\nu}=0$ is assumed to be a priori valid (\cite{Fock}, \cite{Dirac}).

The same procedure can also be applied in the mini-superspace approach. By
considering an FLRW space-time, the continuity equation for $\mathcal{T}%
_{\mu\nu}$ becomes a differential equation that involves $\rho$, $P$ and the
scale factor $a$. Substitution of the equation of state that we mentioned
leads to the well known solution
\begin{equation}  \label{enden}
\rho = m\, a^{-3(1+\gamma)}
\end{equation}
where $m$ is a constant of integration. The addition to the mini-superspace
Lagrangian of an extra term $L_m = -2 \sqrt{-g} \rho = -2 N\, m\, a^{-3
\gamma}$ can be seen that correctly reproduces the set of the reduced
Einstein's equations. In the parametrization in which the potential is constant,
i.e. when we set
\begin{equation}  \label{Nfl}
N = \frac{n }{2 \left(m\, a^{-3 \gamma } -3\, k\, a +a^3 V(\phi)\right)}
\end{equation}
with $n$ being the new ``lapse" function, the aforementioned Lagrangian is
written as
\begin{equation}  \label{Lagfl}
L = \frac{2 a}{ n}\left(m\, a^{-3 \gamma }+a^3 V(\phi )-3\, a\, k\right)
\left( -6\, \dot{a}^2 + \epsilon\, a^2 \dot{\phi}^2 \right) - n .
\end{equation}
It is an easy task to verify that the Euler-Lagrange equations of %
\eqref{Lagfl} are equivalent to Einstein's equations
\begin{equation}  \label{Einfluid}
R_{\mu\nu}- \frac{1}{2} g_{\mu\nu} R = T_{\mu\nu} + \mathcal{T}_{\mu\nu}
\end{equation}
when reduced by the ansatz of an FLRW space-time.

Thus, the corresponding mini-superspace metric we are interested in is
\begin{equation}  \label{minisupfl}
G_{\mu\nu} = 4\, a \left(m\, a^{-3 \gamma }+a^3 V(\phi )-3\, a\, k\right)
\begin{pmatrix}
-6 & 0 \\
0 & \epsilon\, a^2%
\end{pmatrix}%
.
\end{equation}
We aim to consider both cases of a spatially flat and a non-flat universe,
but since the treatment is quite similar to what has already been done we
shall state the results in a more condense manner.

\subsection{The $k=0$ case}

The vector $\xi = \frac{\partial}{\partial \phi}$ remains a conformal
Killing vector satisfying the relation
\begin{equation}
\emph{\textsterling}_\xi G_{\mu\nu} = \frac{a^{3 (\gamma +1)} V^{\prime
}(\phi )}{a^{3 (\gamma +1)} V(\phi )+m} G_{\mu\nu} .
\end{equation}
and the corresponding nonlocal integral of motion is
\begin{equation}  \label{intflflat}
\begin{split}
Q &= p_\phi + \int\!\! n \frac{a^{3 (\gamma +1)} V^{\prime }(\phi )}{a^{3
(\gamma +1)} V(\phi )+m} dt \\
&= \frac{4\, \epsilon\, a^{3(1- \gamma) } \dot{\phi} \left(a^{3 (\gamma +1)}
V(\phi )+m\right)}{n} + \int\!\! n \frac{a^{3 (\gamma +1)} V^{\prime }(\phi )%
}{a^{3 (\gamma +1)} V(\phi )+m} dt
\end{split}%
\end{equation}
It can be seen that $Q=const.$ together with the quadratic constraint
equation $\frac{\partial L}{\partial n}=0$ satisfy the complete set of
Euler-Lagrange equations for the case $k=0$ and thus Einstein's equations.

We proceed in the same manner as previously: at first we fix the gauge by
setting $\phi(t)=t$, which allows us to easily re-parameterize the potential
as a time function. Next, a suitable parametrization of the ``lapse" $n$ can
be chosen so that the equation $Q=const.$ can be turned into a local
expression through writing down the integrand as a total derivative. The
procedure is almost similar to the case $k\neq 0$ without fluid. So, for the
sake of brevity, we skip the full calculations and present the complete
solution of the system in the gauge $\phi=t$:
\begin{align}
a(t) =& e^{\omega/6} \\ \label{sclapsefl}
n(t) =& \left(\frac{6}{\dot{\omega}}-\frac{\dot{\omega}}{\epsilon}\right)
\left(-\frac{1}{2} c_1 e^{I}-2\, \epsilon\, m\, e^{I} \int \frac{e^{-\left(%
\frac{1}{2} (\gamma -1) \omega + I \right)}}{\dot{\omega}} dt + \frac{1}{3}
m\, e^{-\frac{1}{2} (\gamma -1) \omega}\right)^{1/2} \\
\begin{split}
V(t) = & \frac{3\, e^{-\frac{1}{2} (\gamma +2) \omega}}{2\, \dot{\omega}^2} %
\Bigg[ \left(c_1 + 4 \, \epsilon\, m \int \frac{e^{\left(-\frac{1}{2}
(\gamma -1) \omega - I \right)}}{\dot{\omega}} dt \right) \left(6\,
\epsilon- \dot{\omega}^2\right) e^{\left(\frac{1}{2} \gamma \omega + I
\right)} \\
& - 4\, \epsilon\, m\, e^{\frac{\omega }{2}}\Bigg]
\end{split}%
\end{align}
with $c_1$ being a constant of integration, $\omega(t)$ an unspecified non
constant function (reflecting the arbitrariness of the scalar field
potential) and $I$ being given by
\begin{equation}  \label{defIom}
I = \omega - 6\int \!\! \frac{\epsilon}{\dot{\omega}} dt.
\end{equation}
Note here that \eqref{sclapsefl} is the scaled lapse and not the one that
enters the metric. The latter is being given by \eqref{Nfl} (with $k=0$).

As also done in the previous sections, the above expressions can be
significantly simplified when performing a suitable time parametrization. In
this case if one adopts the transformation
\begin{equation}  \label{timetrfl1}
\phi(t)=t= \pm \int\!\! \sqrt{\frac{1}{6\, \epsilon} \left(\frac{\gamma +1}{2}+%
\frac{S^{\prime \prime }(\omega )}{S^{\prime }(\omega )}\right)} d\omega
\end{equation}
with $S(\omega)$ being parameterized as
\begin{equation}  \label{StoF2}
S(\omega) = \exp \left(-6 (\gamma +1) m \int\! e^{F(\omega )-\frac{1}{2}
(\gamma +1) \omega } \, d\omega \right)-\frac{3\, c_1}{(\gamma +1) m},
\end{equation}
the emerging line element takes the exact general form of %
\eqref{lineel1b}; of course, in this case the corresponding potential becomes
\begin{equation}  \label{potfl1}
V(\omega) = \frac{1}{12} e^{-F(\omega )} \left(1-F^{\prime }(\omega )\right)+%
\frac{1}{2} (\gamma -1)\, m\, e^{-\frac{1}{2} (\gamma +1) \omega }
\end{equation}
while the scalar field $\phi(\omega)$ is given, through \eqref{timetrfl1} and %
\eqref{StoF2}, as
\begin{equation}  \label{phi3b}
\phi(\omega) = \pm \int \left[ \frac{1}{6\, \epsilon} \left( F'(\omega) -6 (\gamma +1) m\, e^{F(\omega )- \frac{1}{2} (\gamma +1) \omega }\right) \right]^{1/2} \! d\omega .
\end{equation}
It can be easily checked that relations \eqref{lineel1b}, \eqref{potfl1} and %
\eqref{phi3b} solve Einstein's equation \eqref{Einfluid} with a scalar field
plus a perfect fluid for an arbitrary function $F(\omega)$. As a result,
even in this case, the system has been fully integrated without having to
choose a specific form for the potential.

Once more, with the help of \eqref{efffluid}, one can derive in this gauge
the relations for the energy density $\rho_\phi$ and the pressure $P_\phi$
owed to the scalar field
\begin{subequations}
\begin{align}
\rho_\phi & = \frac{1}{12} e^{-F(\omega )}- m\, e^{-\frac{1}{2} (\gamma +1)
\omega } \\
P_\phi & = \frac{1}{12} e^{-F(\omega )} \left(2 F^{\prime }(\omega
)-1\right) - \gamma \, m\, e^{-\frac{1}{2} (\gamma +1) \omega }.
\end{align}
On the other hand, one can easily verify that for the perfect fluid
\end{subequations}
\begin{equation}  \label{rhofl}
\rho = m\, e^{-\frac{1}{2} (\gamma +1) \omega },
\end{equation}
with the pressure being given of course by $P= \gamma \rho$. The co-moving
velocities for the perfect fluid, $\tilde{u}_\mu$, and the one constructed
by the scalar field, $u_\mu=\frac{\phi_{,\mu}}{\sqrt{-g^{\kappa\lambda}%
\phi_{,\kappa}\phi_{,\lambda}}} = e^{F(\omega)/2}$ are by definition the
same for the given line element (a possible difference in sign is of no
significance since only quadratic expressions appear in the energy momentum
tensor). Thus, one can immediately add the energy densities and pressures to
get the net quantities
\begin{subequations}
\begin{align}
\rho_{\mathrm{tot}} & = \rho_\phi + \rho = \frac{1}{12} e^{-F(\omega )} \\
P_{\mathrm{tot}} & =P_\phi + P = \frac{1}{12} e^{-F(\omega )} \left(2
F^{\prime }(\omega )-1\right),
\end{align}
which are identical to expressions \eqref{rhoP1}. The same result can also
be obtained formally by adding the two energy-momentum tensors. The sum of $%
T_{\mu\nu} + \mathcal{T}_{\mu\nu}$ leads to the same energy-momentum tensor
that is obtained in the $k=0$ case without an additional fluid.

As a result, we can conclude that, as expected in the context of a FLRW
geometry, the perfect fluid can always be ``absorbed" by the scalar field in
the following sense: A system with a perfect fluid (with a linear barotropic
equation) and a (minimally coupled) scalar field \eqref{phi3b} with
potential \eqref{potfl1}, exhibits the same dynamical behaviour in
comparison to another cosmological system possessing a single (minimally
coupled) scalar field \eqref{phi1b} with potential \eqref{potential1b}.

\subsection{The $k \neq$ 0 case}

In this section we conclude our analysis by considering the open/closed
universe cases. The vector $\xi = \frac{\partial}{\partial \phi}$ is once
more a conformal Killing vector, for which
\end{subequations}
\begin{equation}
\emph{\textsterling}_\xi G_{\mu\nu} = \frac{a^{3 \gamma +3} V^{\prime }(\phi
)}{-3 k a^{3 \gamma +1}+a^{3 \gamma +3} V(\phi )+m} G_{\mu\nu}
\end{equation}
holds. The nonlocal integral of motion that corresponds to $\xi$ is
\begin{equation}
\begin{split}
Q = & p_\phi + \int\!\! n(t)\frac{a^{3( \gamma +1)} V^{\prime }(\phi)}{-3 k
a^{3 \gamma +1}+a^{3 (\gamma +1)} V(\phi )+m} dt \\
= & \frac{4\, \epsilon\, a ^{3(1- \gamma) } \dot{\phi} \left(-3 k a^{3
\gamma +1}+a^{3 (\gamma +1)} V(\phi) + m\right)}{n} + \int\!\! \frac{n a^{3
(\gamma +1)} V^{\prime }(\phi)}{-3 k a^{3 \gamma +1}+a^{3 (\gamma +1)}
V(\phi )+m} dt
\end{split}%
\end{equation}
As in all previous cases, relations $Q=const.$ and $\frac{\partial L}{%
\partial n} = 0$ are sufficient to completely integrate the system of the
Euler-Lagrange equations of motion. In the gauge $\phi = t$ the solution
becomes:
\begin{align}
a(t) = & e^{\omega/6} \\
\begin{split}
n(t) = & \frac{\left(\frac{6}{\dot{\omega}}-\frac{\dot{\omega}}{\epsilon}%
\right)}{\sqrt{6}} \Bigg(2\, m\, e^{-\frac{1}{2} (\gamma -1) \omega} -3\,
c_1\, e^{I} \\
& - 12\, \epsilon\, e^{I} \int \frac{e^{-I - \frac{1}{2} (\gamma -1) \omega}
\left(m - 3 k e^{\frac{1}{6} (3 \gamma +1) \omega}\right)}{\dot{\omega}} dt
- 6 k e^{\frac{2 \omega}{3}} \Bigg)^{1/2}
\end{split}
\\
\begin{split}
V(t) = & - \frac{3 e^{-\frac{1}{2} (\gamma +2) \omega}}{2 \dot{\omega}^2} %
\Bigg[ \left(c_1 +4\, \epsilon \int \frac{e^{-I - \frac{1}{2} (\gamma -1)
\omega} \left(m-3 k e^{\frac{1}{6} (3 \gamma +1) \omega}\right)}{\dot{\omega}%
} dt \right) \left(\dot{\omega}^2 - 6\, \epsilon\right) e^{I + \frac{1}{2}
\gamma \omega} \\
& - 12 \, k\, e^{\frac{1}{6} (3 \gamma +4) \omega}+4\, m\, e^{\frac{\omega}{2%
}}\Bigg]
\end{split}%
\end{align}
with $c_1$ being a constant of integration, $I$ given again by \eqref{defIom}
and $\omega(t)$ remaining an arbitrary non-constant function.

By performing a time transformation and introducing a new function $%
S(\omega) $
\begin{equation}  \label{timetrfl2}
\phi(\omega)=t(\omega)=\pm \int\!\! \left(\frac{-18 S^{\prime \prime \prime
}(\omega) + 3 (3 \gamma -1) S^{\prime \prime }(\omega )+(3 \gamma +1)
S^{\prime }(\omega )}{18\,\epsilon\, \left((3 \gamma +1) S^{\prime }(\omega
)-6 S^{\prime \prime }(\omega )\right)}\right)^{1/2} d\omega ,
\end{equation}
which we can associate to an another function $F(\omega)$ through the relation
\begin{equation}  \label{StoF3}
\begin{split}
S^{\prime \prime }(\omega) = & \frac{1}{6} e^{-\left(\gamma +\frac{7}{6}%
\right) \omega } \left(18\, c_1\, e^{\left(\gamma +\frac{5}{6}\right) \omega
+F(\omega)} \right. \\
& \left. +S^{\prime}(\omega)\left((3 \gamma +1) e^{\left(\gamma +\frac{7}{6}\right)
\omega }+72 k e^{\left(\gamma +\frac{5}{6}\right) \omega +F(\omega )}-36
(\gamma +1)\, m\, e^{\frac{1}{6} (3 \gamma +4) \omega +F(\omega )}\right)
\right. \\
& \left. - 12\, (3 \gamma +1)\, k\, S(\omega ) e^{\left(\gamma +\frac{5}{6}%
\right) \omega +F(\omega )}\right)
\end{split}%
\end{equation}
we are led to line element \eqref{lineel2b}. The potential is parameterized
with respect to another function $F(\omega)$ as
\begin{equation}  \label{potfl2}
V(\omega) = \frac{1}{12} e^{-F(\omega )} \left(1-F^{\prime }(\omega )\right)
+2\, k\, e^{-\omega/3} + \frac{1}{2} (\gamma -1)\, m\, e^{-\frac{1}{2}
(\gamma +1) \omega }
\end{equation}
and together with the aforementioned metric \eqref{lineel2b} and
\begin{equation}\label{phi4b}
  \begin{split}
    \phi(\omega) = \pm \int\! & \left[  \frac{1}{6\, \epsilon} \left(  F'(\omega)+ 12\, k \, e^{-\omega/3 +F(\omega )} \right. \right. \\
      & \left.  \left. -6\, (\gamma +1)\, m\, e^{F(\omega )-\frac{1}{2} (\gamma +1) \omega } \right) \right]^{1/2}\! d\omega
  \end{split}
\end{equation}
solve the Einstein's equations for a minimally coupled scalar field in the
presence of the perfect fluid that we considered. As in all previous cases
the function $F(\omega)$ remains free, since we have not chosen a particular
scalar field potential.

Of course, the relations regarding the energy density and the pressure of
the scalar field can also be derived as functions of $\omega$, being
\begin{subequations}
\begin{align}
\rho_\phi = & \frac{1}{12} e^{-F(\omega )}+ 3\, k\, e^{-\frac{\omega }{3}} -
m \, e^{-\frac{1}{2} (\gamma +1) \omega }, \\
P_\phi = & \frac{1}{12} e^{-F(\omega )} \left(2 F^{\prime }(\omega
)-1\right)- k \,e^{-\omega/3} -m \, \gamma\, e^{-\frac{1}{2} (\gamma +1)
\omega }
\end{align}
respectively.

Again, the energy density and pressure of the fluid are given by %
\eqref{rhofl} and $P=\gamma \rho$. As in the previous case - and for the
same reason - the total quantities $\rho_{\mathrm{tot}}$ and $P_{\mathrm{tot}%
}$ can be retrieved by a simple addition
\end{subequations}
\begin{subequations}
\begin{align}
\rho_{\mathrm{tot}} & =\rho_\phi+ \rho = \frac{1}{12} e^{-F(\omega )}+ 3\,
k\, e^{-\frac{\omega }{3}} \\
P_{\mathrm{tot}} & =P_\phi + P = \frac{1}{12} e^{-F(\omega )} \left(2
F^{\prime }(\omega )-1\right)- k \,e^{-\omega/3}
\end{align}
and are identical to expressions \eqref{rhoP2} obtained in the case $k\neq 0$
of a single scalar field. The same is also true for the total energy
momentum tensor $T_{\mu\nu}+\mathcal{T}_{\mu\nu}$. Henceforth, the statement
that we made in the case $k=0$, also applies here. A system that possesses
both a scalar field and a perfect fluid can be simulated by another with
just a single appropriate scalar field.

In addition, we can consider an arbitrary number, say $\nu \in \mathbb{N}$, of perfect fluids
each one satisfying an equation of state of the form
\end{subequations}
\begin{equation}
P_{i}=\gamma _{i}\rho _{i},\quad \quad i=1,...,\nu.
\end{equation}%
The Einstein plus Klein-Gordon system of equations in this case
\begin{subequations}
\label{systnfluid}
\begin{align}
& R_{\mu \nu }-\frac{1}{2}g_{\mu \nu }R=T_{\mu \nu }+\sum_{i=1}^{\nu}\mathcal{T}_{\mu
\nu }^{(i)} \\
& \epsilon \Box \phi (\omega )+\frac{1}{\phi ^{\prime }(\omega )}V^{\prime
}(\omega )=0,
\end{align}
\end{subequations}
where $\mathcal{T}_{\mu \nu }^{(i)}$ is the energy momentum tensor of the $i$ labeled
fluid, have a solution given by line element \eqref{lineel2b} with the scalar
field being
\begin{equation}
  \begin{split}
    \phi(\omega) =\pm  \int\! & \left[  \frac{1}{6\, \epsilon} \left(  F'(\omega)+ 12\, k \, e^{-\omega/3 +F(\omega )} \right. \right. \\
      & \left.  \left. -6\sum_{i=1}^{\nu} (\gamma_i +1)\, m_i\, e^{F(\omega )-\frac{1}{2} (\gamma_i +1) \omega } \right) \right]^{1/2}\! d\omega
  \end{split}
\end{equation}
while the corresponding potential is
\begin{equation}
V(\omega )=\frac{1}{12}e^{-F(\omega )}\left( 1-F^{\prime }(\omega )\right)
+2\,k\,e^{-\omega /3}+\frac{1}{2}\sum_{i=1}^{\nu}(\gamma _{i}-1)\,m_{i}\,e^{-%
\frac{1}{2}(\gamma _{i}+1)\omega }.
\end{equation}%
The energy density of each fluid is of course given by
\begin{equation}
\rho _{i}=m_{i}e^{-\frac{1}{2}(\gamma _{I}+1)\omega },
\end{equation}%
with the $m_{i}$'s being constants of integration. Finally, the expressions
for the energy density and pressure of the effective fluid that simulates
the behaviour of the scalar field are
\begin{subequations}
\begin{align}
\rho _{\phi }=& \frac{1}{12}e^{-F(\omega )}+3\,k\,e^{-\frac{\omega }{3}%
}-\sum_{i=1}^{\nu}m_{i}\,e^{-\frac{1}{2}(\gamma _{i}+1)\omega }, \\
P_{\phi }=& \frac{1}{12}e^{-F(\omega )}\left( 2F^{\prime }(\omega )-1\right)
-k\,e^{-\omega /3}-\sum_{i=1}^{\nu}m_{i}\,\gamma _{i}\,e^{-\frac{1}{2}(\gamma
_{i}+1)\omega }.
\end{align}%
It is straightforward to check that the previous relations identically
satisfy the system of equations \eqref{systnfluid}. Thus, completing the
solution of a general scalar field with in the presence of an arbitrary
number of perfect fluids.

\section{Particular Solutions\label{examples}}

In this section, in order to demonstrate the power and applicability of our results, we show how specific
models can be studied by putting in use the equations of state %
\eqref{eqofst1} and \eqref{eqofst2}, depending on whether we are in the
spatially flat case or not. We do that for three special forms for the
equation of state parameter of the total fluid in which our cosmological
model is that of quintessence scalar field without matter source. The cases
that we study are: (a) constant equation of state parameter for vanishing
and nonvanishing spatial curvature, (b) exponentially dependent parametric
dark energy model, and (c) the logarithmic dark energy model.

\subsection{A constant equation of state parameter}

Let us first consider the $k=0$ case: thus, due to \eqref{rhoP1}, we demand that
\end{subequations}
\begin{equation}
\frac{P}{\rho }=2\,F^{\prime }(\omega )-1=\gamma
\end{equation}%
with $\gamma $ being a constant different from $-1$ (since in our definition
of $F$ we needed it to be a non-constant function). Hence, we are led to the
solution
\begin{equation}
F(\omega )=\frac{1}{2}(\gamma +1)\omega
\end{equation}%
where we have omitted the integration constant, because - as can be seen
from the induced line element \eqref{lineel1b} - it is not essential and can
be absorbed with an appropriate coordinate transformation. The corresponding
scalar field and potential are given by (in what follows we choose to work
just with the plus solution of every $\phi (\omega )$)
\begin{equation}
\phi (\omega )=\frac{1}{2}\sqrt{\frac{1+\gamma }{3\, \epsilon }}\omega ,\quad
\quad V(\phi (\omega ))=-\frac{1}{24}(\gamma -1)e^{-\frac{1}{2}(\gamma
+1)\omega },
\end{equation}%
which imply the exponential relation $V(\phi )=-\frac{1}{24}(\gamma
-1)e^{\epsilon \sqrt{3\epsilon (\gamma +1)}\phi }$. We can go over to the
gauge where the lapse function of the metric is one, i.e. we shall adopt the
cosmological time variable. The transformation we need is
\begin{equation}
\int e^{F(\omega )/2}d\omega =\tau \Rightarrow \omega =\frac{4}{\gamma +1}\ln \left(
\frac{1}{4}(\gamma +1)\tau \right)
\end{equation}%
and in this gauge the corresponding solution becomes the well known power
law for the scale factor and the logarithm relation for the scalar field
\cite{russo2004exact}
\begin{equation}
a(\omega )=e^{\omega /6}\Rightarrow a(\tau )\propto \tau ^{\frac{2}{3(\gamma
+1)}},\quad \quad \phi (\tau )=\frac{2\ln \left( \frac{1}{4}(\gamma +1)\tau
\right) }{\sqrt{3\,\epsilon (\gamma +1)}}.
\end{equation}

The situation is slightly more complicated in the $k\neq 0$ case. The
differential equation at hand is
\begin{equation} \label{eqstex1}
  \frac{P}{\rho} = \gamma \Rightarrow (\gamma +1) e^{\omega /3}-2\, e^{\omega /3} F'(\omega )+12 (3 \gamma +1) k e^{F(\omega )} =0,
\end{equation}
where now of course $\rho $ and $P$ are given by relations \eqref{rhoP2}.

First of all, let us check the specific case when $\gamma =-\frac{1}{3}$.
The solution of \eqref{eqstex1} is
\begin{equation}
F(\omega )=\frac{\omega }{3},  \label{solconstgamma0}
\end{equation}%
where once more the constant of integration has been assumed zero, since it
is not essential for the geometry, as one can see by line element %
\eqref{lineel2b}. This solution for $\gamma =-\frac{1}{3}$ is
distinguished from all other values of $\gamma $, for which the
corresponding models depend on two parameters as we shall immediately see.

When $\gamma \neq -\frac{1}{3}$, the general solution of \eqref{eqstex1} is
\begin{equation}
F(\omega )=\frac{1}{3}\left[ \omega -3\log \left( e^{\frac{1}{6}(3\gamma
+1)(\mu -\omega )}-36\,k\right) \right] ,  \label{solconstgamma}
\end{equation}%
where $\mu $ is the integration constant, which unlike the previous case, is
essential for line element \eqref{lineel2b}. The scalar field and potential
become (in what follows - for simplicity - we choose to express the results
for the standard scalar field, i.e. we set $\epsilon =+1$)
\begin{align}
\phi (\omega )& =-\frac{2\sqrt{3}\sqrt{\gamma +1}\ln \left( e^{\frac{1}{12}%
(3\gamma +1)(\mu -\omega )}+\sqrt{e^{\frac{1}{6}(3\gamma +1)(\mu -\omega
)}-36k}\right) }{3\gamma +1} \\
V(\omega )& =-\frac{1}{24}(\gamma -1)e^{\frac{1}{6}(\mu (3\gamma
+1)-3(\gamma +1)\omega )},
\end{align}%
while the line element can be written as
\begin{equation}
ds^{2}=-\frac{e^{\omega /3}}{e^{\frac{1}{6}(3\gamma +1)(\mu -\omega )}-36\,k}%
d\omega ^{2}+\frac{e^{\omega /3}}{1-kr^{2}}dr^{2}+r^{2}(d\theta ^{2}+\sin
^{2}\theta d\phi ^{2}).
\end{equation}%
It is clear by the form of the above metric that, had we tried to express the result in the cosmological time gauge, the solution would not be possible to be given in terms of elementary functions, since
\begin{equation}
\begin{split}
& \int \left( \frac{e^{\omega /3}}{e^{\frac{1}{6}(3\gamma +1)(\mu -\omega
)}-36k}\right) ^{1/2}d\omega =\tau \Rightarrow \\
& \tau =\frac{e^{\omega /6}}{\sqrt{-k}}\,_{2}F_{1}\left( \frac{1}{2},\frac{1%
}{-3\gamma -1};\frac{3\gamma }{3\gamma +1};\frac{e^{\frac{1}{6}(3\gamma
+1)(\mu -\omega )}}{36k}\right) .
\end{split}%
\end{equation}%
where $_{2}F_{1}(a,b;c;x)$ is the Gauss Hypergeometric function. Note here,
that the $-k$ in the square root does not restrict $k$ to being $-1$ for
this transformation to be valid, since there exist values of the parameters $%
\mu $, $\gamma $ and $\omega $, for which $_{2}F_{1}$ can be purely
imaginary.

However, useful results can be extracted in the gauge where time is $\omega $%
, if the various cosmological parameters are expressed in parametric form.
It is true that the Hubble $H$, deceleration $q$ and jerk $j$ parameters can
be given in an arbitrary gauge $N(t)$ by
\begin{subequations}
\label{cosmpar}
\begin{align}
H(t)& =\frac{1}{a\,N}\frac{da}{dt} \\
q(t)& =-a\left( \frac{1}{N}\frac{da}{dt}\right) ^{-2}\frac{1}{N}\frac{d}{dt}%
\left( \frac{1}{N}\frac{da}{dt}\right) \\
j(t)& =\frac{1}{a\,H(t)^{3}\,N}\frac{d}{dt}\left( \frac{1}{N}\frac{d}{dt}%
\left( \frac{1}{N}\frac{da}{dt}\right) \right) .
\end{align}

With the time variable being $\omega $ we can easily derive all the previous
parameters as functions of the scale factor $a=e^{\omega /6}$, by simply
setting in the expressions $\omega =6\ln a$. Of course, the gauge function
is given by $N(\omega )=e^{F(\omega)/2}$, where $\omega = 6 \ln a $, and $F\left( \omega \right) $ is taken
to be \eqref{solconstgamma}.

The corresponding relations are
\end{subequations}
\begin{subequations}
\begin{align}
H(a)& =\sqrt{\frac{e^{\frac{1}{6}(3\gamma +1)\mu }}{36a^{3(\gamma +1)}}-%
\frac{k}{a^{2}}}  \label{hb.001} \\
q(a)& =\frac{(3\gamma +1)e^{\frac{1}{6}(3\gamma +1)\mu }}{2\left( e^{\frac{1%
}{6}(3\gamma +1)\mu }-36ka^{3\gamma +1}\right) }  \label{hb.002} \\
j(a)& =\frac{\left( 9\gamma ^{2}+9\gamma +2\right) e^{\frac{1}{6}(3\gamma
+1)\mu }}{2\left( e^{\frac{1}{6}(3\gamma +1)\mu }-36ka^{3\gamma +1}\right) }
\label{hb.003}
\end{align}%
and their behaviour with respect to the scale factor $a$ and according to
various values of the essential constants $\gamma $ and $\mu $ can be easily
derived.

In the present epoch, in which by convention we take it to be $a=1$, we deduce from (\ref{hb.001}) that the integration
constant $\mu $ is related to the Hubble constant; specifically, we find that
\end{subequations}
\begin{equation}
\mu =\frac{12}{3\gamma +1}\ln \left[ 6\left( H_{0}+k\right) \right], \quad\quad \gamma
\neq -\frac{1}{3}.
\end{equation}

For $\gamma =-\frac{1}{3}$, the function $H(a)$ can be calculated by %
\eqref{cosmpar} with the use of solution \eqref{solconstgamma0}, again under
the substitution $\omega =6\ln a$. In this particular case one can see that $%
H(a)=\frac{1}{6\,a}$, while $q(a)=j(a)=0$.

\subsection{An exponential equation of state parameter\label{secondex}}

\begin{figure}[h]
\begin{subfigure}{.5\textwidth}
  \centering
  \includegraphics[scale=0.6]{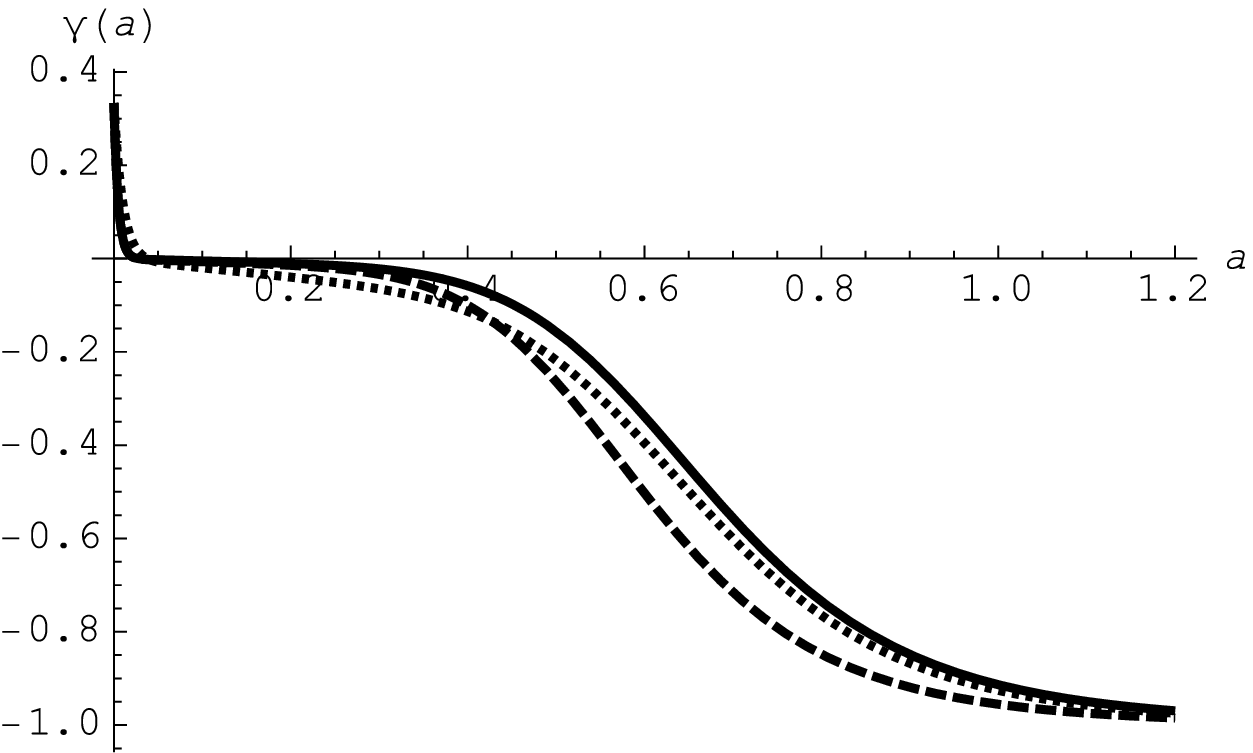}
  \caption{EoS parameter $\gamma(a)$}
  \label{figC3a}
    \end{subfigure}%
\begin{subfigure}{.5\textwidth}
  \centering
  \includegraphics[scale=0.6]{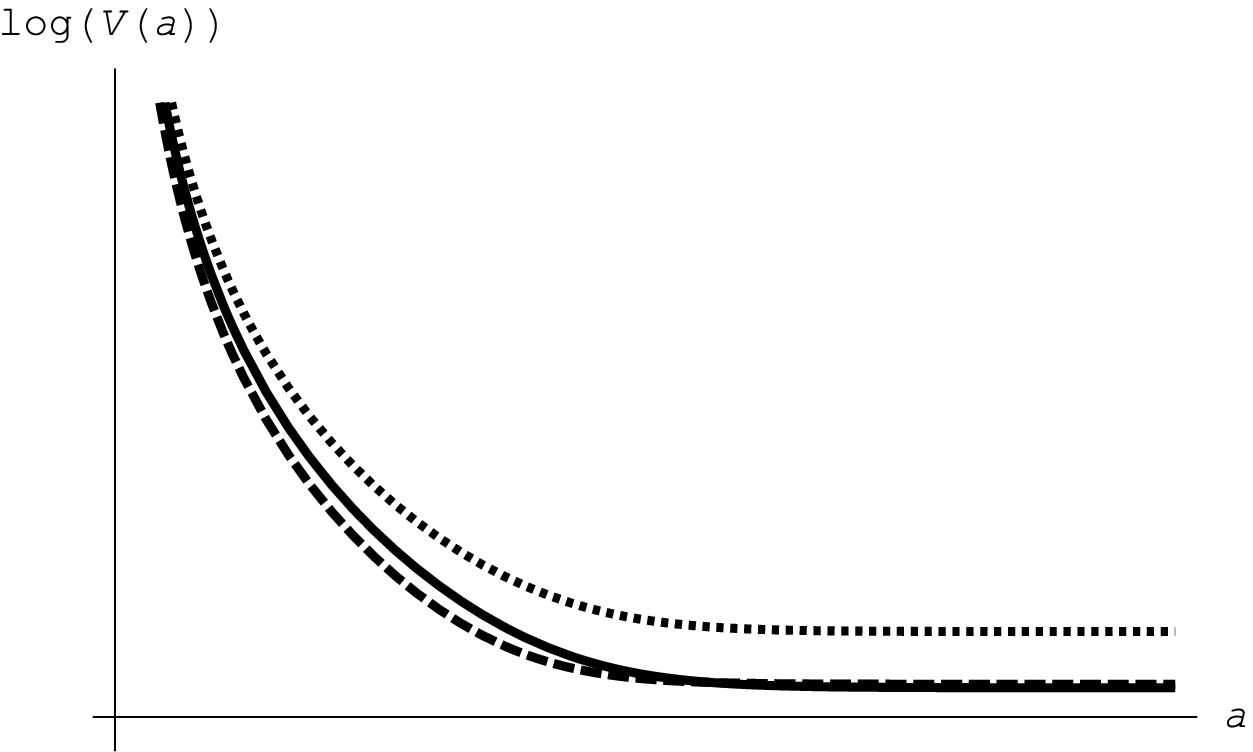}
  \caption{Potential $V(a)$}
  \label{figC3b}
  \end{subfigure}
\caption{The qualitative behavior of the equation of state parameter (EoS) $%
\protect\gamma (a)$ and of the potential $V(a)$ are given in figures \protect
\ref{figC3a} and \protect\ref{figC3b} respectively. The solid lines are for
the constants $\left( \protect\lambda ,\protect\mu ,\protect\nu \right)
=\left( 1/10,1/20,200\right) ,~$the dash-dash lines for $\left( \protect%
\lambda ,\protect\mu ,\protect\nu \right) =\left( 1/20,1/15,200\right) $,
while the dot-dot lines are for the constants $\left( \protect\lambda ,%
\protect\mu ,\protect\nu \right) =\left( 1/10,1/5,100\right) $.}
\label{figexp1}
\end{figure}

As a second example we consider that the equation of state parameter for the
total fluid has the following form%
\begin{equation}
\gamma \left( a\right) =-1+\frac{\lambda }{\lambda +a^{\sigma }}e^{-\mu a}+%
\frac{1}{3}e^{-\nu a}\text{,}  \label{pex.01}
\end{equation}%
the reason being that as the current experimental data suggest, in the early universe we must have $\gamma
\left( a\rightarrow 0\right) \simeq \frac{1}{3}$, while in the late universe
$\gamma \left( a\rightarrow 1\right) \simeq -1$ for positive values $\sigma ,\mu ,\nu
$. However there will be an epoch in which the equation of state parameter
will have a linear behavior around the $\gamma =0$. For simplicity in the
following we consider that $\sigma =6$, where the evolution of (\ref{pex.01}%
) is given in figure \ref{figC3a}.

Furthermore from \eqref{eqofst1}, when we assume that the scalar field has
the behavior (\ref{pex.01}) we find that.

\begin{equation}
F(\omega )=\frac{1}{2}\lambda \int \!\!\frac{\exp \left( -\mu e^{\omega
/6}\right) }{\lambda +e^{\omega }}\,d\omega +\mathrm{Ei}\left( -\nu
e^{\omega /6}\right) ,  \label{pex.02}
\end{equation}%
with $\mathrm{Ei}(x)=-\int_{-x}^{+\infty }\frac{e^{-s}}{s}ds$ being the exponent integral function. The
integration constant in $F(\omega )$ has been set to zero.

In the case of a quintessence scalar field, from equation (\ref{phi1b}), we have that
\begin{subequations}
\begin{equation}
\phi (a)=\pm \int \!\frac{1}{a}\left( \frac{3\,\lambda \,e^{-\mu \,a}}{%
a^{6}+\lambda }+e^{-\nu \,a}\right) ^{1/2}\!\!da,
\end{equation}%
while for the potential we have that
\end{subequations}
\begin{equation}
V(a)=\frac{1}{72}\left( -\frac{3\lambda e^{-a\mu }}{a^{6}+\lambda }-e^{-a\nu
}+6\right) \exp \left[ -\mathrm{Ei}\left( -\nu \,a\right) -3\lambda \int
\frac{e^{-\mu \,a}}{a\left( a^{6}+\lambda \right) }\,da\right] ,
\end{equation}%
where in figure \ref{figC3b} the evolution of $V\left( a\right) $ is given.
Here we would like to remark that if we had considered that there were also
extra fluid terms, then the solutions for the scalar
field, i.e. $\phi \left( a\right) $,~$V\left( a\right) $, would be
different. But in any case the solution (\ref{pex.02}) would be the same if
we assume that (\ref{pex.01}) describes the equation of state parameter for
the total fluid.

\begin{figure}[h]
\begin{subfigure}{.45\textwidth}
  \centering
  \includegraphics[scale=0.6]{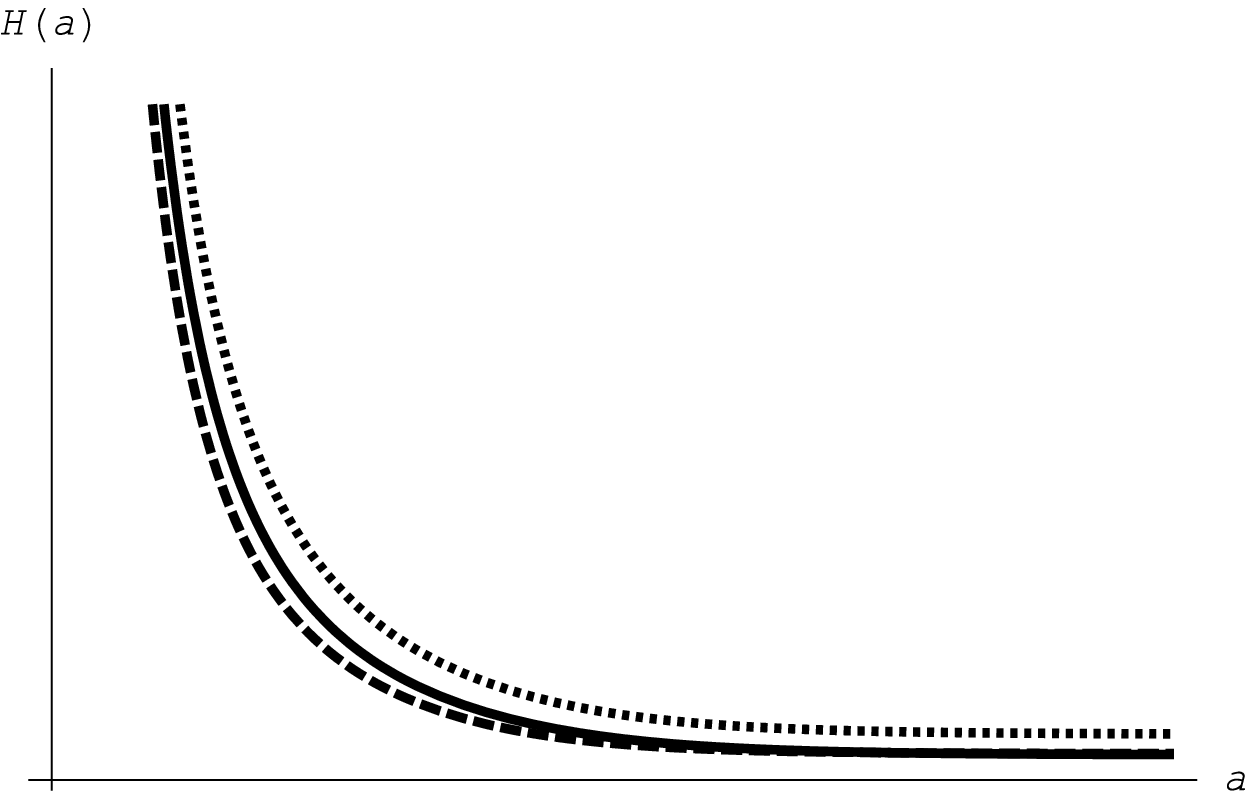}
  \caption{Hubble parameter $H(a)$}
  \label{figC3c}
  \end{subfigure}
\begin{subfigure}{.45\textwidth}
  \centering
  \includegraphics[scale=0.6]{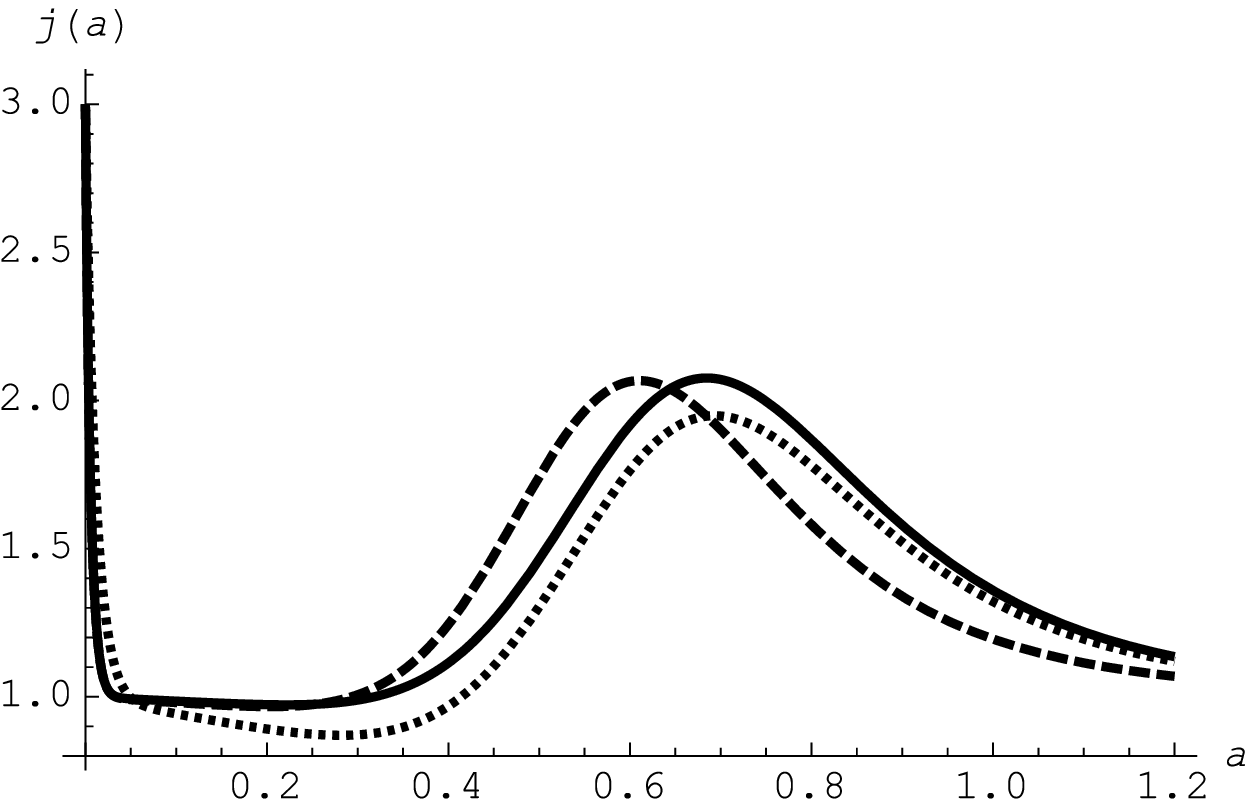}
  \caption{Jerk parameter $j(a)$}
  \label{figC3d}
  \end{subfigure}
\caption{The qualitative behavior cosmological parameter $H\left( a\right) $%
, and $j\left( a\right) $ are given in figures \protect\ref{figC3c} and
\protect\ref{figC3d} respectively. The solid lines are for the constants $%
\left( \protect\lambda ,\protect\mu ,\protect\nu \right) =\left(
1/10,1/20,200\right) ,~$the dash-dash lines for $\left( \protect\lambda ,%
\protect\mu ,\protect\nu \right) =\left( 1/20,1/15,200\right) $, while the
dot-dot lines are for the constants $\left( \protect\lambda ,\protect\mu ,%
\protect\nu \right) =\left( 1/10,1/5,100\right) $.}
\label{figexp2}
\end{figure}

Finally the Hubble function and the jerk parameters are given as follows
while the qualitative evolution is given in figure \ref{figexp2}.

\begin{equation}
H(a)=\frac{1}{6}\left[ \exp \left( \mathrm{Ei}\left( -\nu \,a\right)
+3\lambda \int \frac{e^{-a\mu }}{a\left( a^{6}+\lambda \right) }\,da\right) %
\right] ^{-1/2},
\end{equation}

\begin{equation}
\begin{split}
j(a)=& \frac{e^{-2(\mu +\nu )a}}{2\left( a^{6}+\lambda \right) ^{2}}\left[
\nu \,a^{13}e^{(2\mu +\nu )a}+a^{12}e^{2\,\mu \,a}\left( 2e^{2\,\nu
\,a}-3e^{\nu \,a}+1\right) \right. \\
& \left. +\lambda \,a^{7}\left( 3\,\mu \,e^{(\mu +2\nu )a}+2\,\nu \,e^{(2\mu
+\nu )a}\right) \right. \\
& \left. +\lambda \,a^{6}\left( 4e^{2(\mu +\nu )a}-6e^{(2\mu +\nu
)a}+9e^{(\mu +2\nu )a}+2e^{2\,\mu \,a}\right) +6\,\lambda \left(
a^{6}+\lambda \right) e^{(\mu +\nu )a}\right. \\
& \left. +\lambda ^{2}a\left( 3\,\mu \,e^{(\mu +2\nu )a}+\nu \,e^{(2\mu +\nu
)a}\right) \right. \\
& \left. +\lambda ^{2}\left( 2e^{2(\mu +\nu )a}-3e^{(2\mu +\nu )a}-9e^{(\mu
+2\nu )a}+e^{2\,\mu \,a}+9e^{2\,\nu \,a}\right) \right] .
\end{split}%
\end{equation}

\subsection{Logarithmic parametric dark energy model}

We consider the logarithmic parametric model \cite{logpar}%
\begin{equation}
\gamma \left( a\right) =\gamma _{0}-\gamma _{1}\ln \left( a\right) ,
\label{model22}
\end{equation}%
which gives that the unknown function in the line element should be

\begin{equation}
F(\omega )=\frac{1}{24}\omega (12\gamma _{0}-\gamma _{1}\omega +12)
\end{equation}%
wherefrom we have that the scalar field and cosmological parameters of the
model are as follows
\begin{subequations}
\begin{align}
\phi (a)=& \mp \frac{2(-\gamma _{1}\frac{\omega }{6}+\gamma _{0}+1)^{3/2}}{%
\sqrt{3}\gamma _{1}}+c_{1} \\
V(a)=& \frac{1}{24}a^{\frac{3}{2}\gamma _{1}\ln (a)-3(\gamma _{0}+1)}(\gamma
_{1}\ln (a)-\gamma _{0}+1)
\end{align}
\end{subequations}
\begin{subequations}
\begin{align}
H(a)& =\frac{1}{6}\left( a^{-\frac{3}{2}\gamma _{1}\ln (a)+3(\gamma
_{0}+1)}\right) ^{-1/2} \\
j(a)& =\frac{1}{2}\big[9\gamma _{1}\ln (a)\left( \gamma _{1}\ln (a)-2\gamma
_{0}-1\right) +9\gamma _{0}(\gamma _{0}+1)+3\gamma _{1}+2\big]
\end{align}

Finally for $c_{1}=0$, we have that the functional form of the potential is
\end{subequations}
\begin{equation}
V(\phi )=\frac{4-6^{1/3}\gamma _{1}^{2/3}\phi ^{2/3}}{48}\exp \left[ \frac{3\left( 6^{2/3}\gamma _{1}^{4/3}\phi
^{4/3}-4\gamma _{0}(\gamma _{0}+2)-4\right) }{8\gamma _{1}}\right].
\end{equation}

\begin{figure}[h!]
\begin{subfigure}{.45\textwidth}
  \centering
  \includegraphics[scale=0.6]{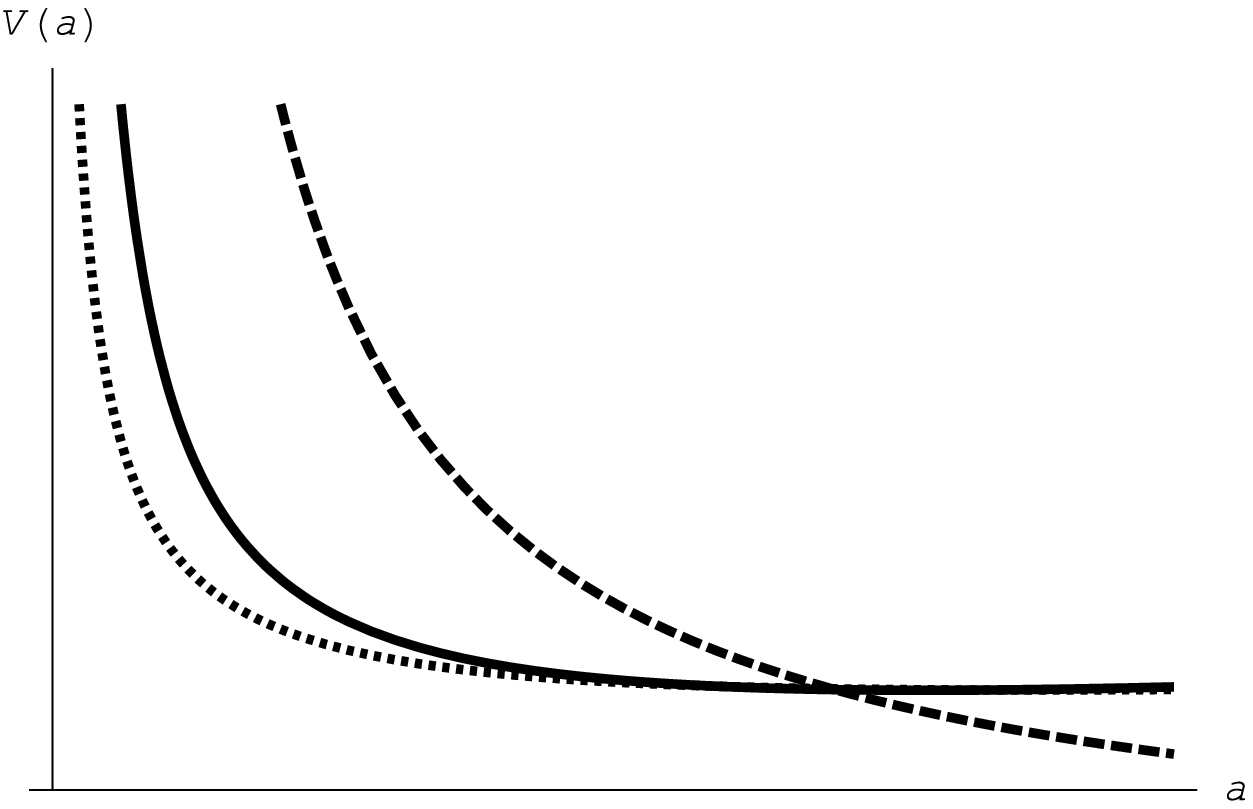}
  \caption{Potential $V(a)$}
  \label{figL3b}
  \end{subfigure}
\begin{subfigure}{.45\textwidth}
  \centering
  \includegraphics[scale=0.6]{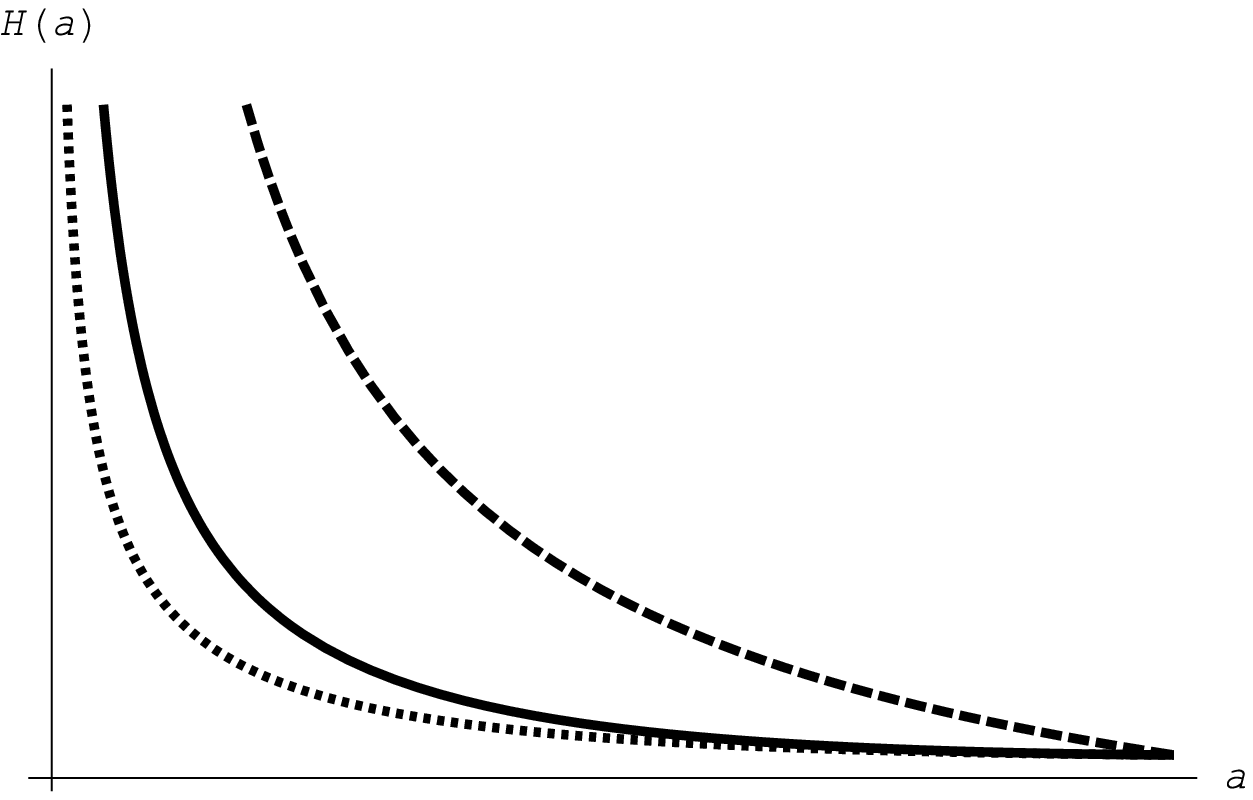}
    \caption{Hubble parameter $H(a)$}
  \label{figL3c}
  \end{subfigure}\\[1ex]
\begin{subfigure}{\linewidth}
  \centering
  \includegraphics[scale=0.6]{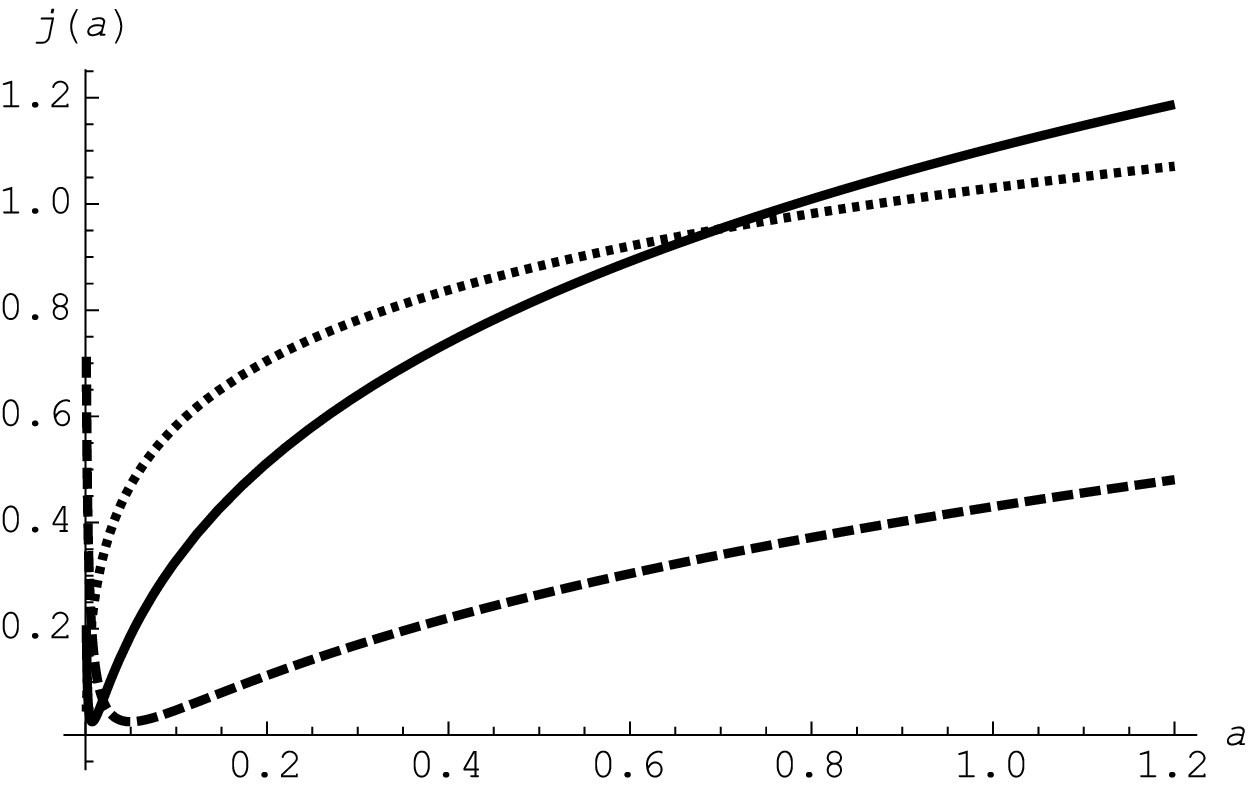}
  \caption{Jerk parameter $j(a)$}
  \label{figL3d}
  \end{subfigure}
\caption{The qualitative evolution of the potential function $V\left(
a\right)$ (figure \protect\ref{figL3b}), for the Hubble function (figure
\protect\ref{figL3c} ) and the jerk parameter (figure \protect\ref{figL3d})
for the logarithmic dark energy model. The solid lines are for the
parameters $\left( \protect\gamma _{0},\protect\gamma _{1}\right) =\left(
-0.99,0.1\right) $,~the dash-dash lines are for $\left( \protect\gamma _{0},%
\protect\gamma _{1}\right) =\left( -0.9,0.1\right) $, and the dot lines are
for the constants $\left( \protect\gamma _{0},\protect\gamma _{1}\right)
=\left( -0.99,0.05\right) $.}
\label{model2a}
\end{figure}

In figures \ref{figL3b}, \ref{figL3c} and \ref{figL3d}, the evolution of the equation of
state parameter, of the scalar field potential~$V\left( \phi \right) $ and
of the cosmological parameters $H\left( a\right) ,~j\left( a\right) $, is
given.

\section{Conclusions}

\label{conclusions}

In this paper we were able, in the context of a FLRW geometry, to derive the general solution for a scalar field, with an arbitrary potential, minimally coupled to Einstein's gravity with or without including perfect fluids. This was made possible mainly by exploiting the reparameterization invariance inherent in constrained systems characterized by \eqref{general_lag}.

The prescription we used was the following: A mini-superspace Lagrangian was constructed in each case we considered. The existence of the Hamiltonian constraint allowed us to define nonlocal integrals of motion corresponding to the conformal symmetries of the mini-superspace. One of these infinite (for a two dimensional configuration space) conserved quantities, together with the quadratic constraint provided us with enough equations to solve the system. In order to turn the nonlocal expression into a first order differential equation, we adopted an appropriate gauge and performed specific reparametrizations so as to succeed in integrating both the first integral and the quadratic constraint. As a result, we can now state that the solution we presented engulfs all possible cosmological configurations regarding a minimally coupled scalar field (apart from the specific case $\phi=$const. that we had to exclude from the analysis, although of course it can be treated in a similar manner).

It could be argued that a general solution of this form, as obtained in an arbitrary gauge, might not be of a major physical importance; mainly due to the fact that the inversion of the function $F(\omega)$, which is essential for expressing the potential as a function of the scalar field, may be transcendental and thus of not of particular use. However, as we demonstrated in the examples, it is a relatively easy task to derive the general expressions regarding the effective perfect fluid related to the scalar field. From this point, the association of a physical behaviour for a given equation of state, written in parametric form, is just a matter of solving a first order ordinary differential equation (or simply an integration with respect to $\omega$ in the spatially flat case without fluids). Additionally, from a mathematical perspective, it is an interesting fact that this two dimensional mini-superspace under consideration is integrable for every well behaved function $V(\phi)$ and an analytic solution can be derived without the need to impose any restrictions on the potential.

Moreover, the solutions we obtained in this work can also prove to be a useful tool for other gravitational configurations. There are several types of theories whose action, under specific transformations, can be mapped to the minimally coupled scalar field of general relativity. For instance we can mention $f(R)$ or several scalar-tensor theories of gravitation (like Brans-Dicke cosmology). The transformations that one uses to go over from the Jordan to the Einstein frame are well known in the literature and we restrain from presenting them here. This link between these theories can be used at any point to transform results from one frame to the other. Nevertheless, we have to note that in the presence of fluids, someone should be careful when making this transition. While we have considered that the fluid terms are not interacting with the scalar field, this property is lost under a conformal transformations and interactions among the former and the scalar field shall arise.

For the completeness of the method that we presented here, we plan - in a forthcoming work - to extend it on applications to cosmological models in scalar-tensor theory with perfect fluids which are not interacting with the scalar field. The possible derivation of analytic solutions, among the two different theories/frames, is important in order to understand better the differences between them.

%\newpage

\begin{acknowledgments}
N. D. acknowledges financial support by FONDECYT postdoctoral grant no.
3150016. A. P. acknowledges financial support by FONDECYT postdoctoral grant
no. 3160121.
\end{acknowledgments}

\bibliography{references}

\end{document}